\documentstyle[epsfig]{elsart}
\begin{document}

\newcommand{\mathbold}[1]{\mbox{\rm\bf #1}}
\newcommand{\mrm}[1]{\mbox{\rm #1}}
\newcommand{\N}{{\cal N}}
\newcommand{\bla}{\hspace{1cm}}
\newcommand{\be}{\begin{equation}}
\newcommand{\ee}{\end{equation}}
\newcommand{\nn}{\nonumber}
\newcommand{\bea}{\begin{eqnarray}}
\newcommand{\eea}{\end{eqnarray}}
\newcommand{\eq}[1]{eq.~(\ref{#1})}
\newcommand{\rfn}[1]{(\ref{#1})}
\newcommand{\Eq}[1]{Eq.~(\ref{#1})}
\newcommand{\gsim}{\ \rlap{\raise 2pt\hbox{$>$}}{\lower 2pt \hbox{$\sim$}}\ }
\newcommand{\lsim}{\ \rlap{\raise 2pt\hbox{$<$}}{\lower 2pt \hbox{$\sim$}}\ }
\newcommand{\ep}{\epsilon}
\newcommand{\D}{\Delta}

\newcommand{\ssu}{$SU(2)_L\times SU(2)_R\times U(1)_{B-L}\,$}
\newcommand{\sul}{$SU(2)_L$}
\newcommand{\sulu}{$SU(2)_L\times U(1)_Y$}
\newcommand{\sur}{$SU(2)_R$}
\newcommand{\thh}{\displaystyle\frac{1}{3}}
\newcommand{\matr}{\left( \begin{array}}
\newcommand{\ematr}{\end{array} \right)}
\newcommand{\g}{\gamma}
\newcommand{\lr}{{left-right symmetric model }}

\newcommand{\ea}{{ et al.}}
\newcommand{\ib}{{\it ibid.\ }}

\newcommand{\np}[1]{Nucl. Phys. {\bf #1}}
\newcommand{\pl}[1]{Phys. Lett. {\bf #1}}
\newcommand{\pr}[1]{Phys. Rev. {\bf #1}}
\newcommand{\prl}[1]{Phys. Rev. Lett. {\bf #1}}
\newcommand{\zp}[1]{Z. Phys. {\bf #1}}
\newcommand{\prep}[1]{Phys. Rep. {\bf #1}}
\newcommand{\rmp}[1]{Rev. Mod. Phys. {\bf #1}}    
\newcommand{\ijmp}[1]{Int. Jour. Mod. Phys. {\bf #1}}
\newcommand{\mpl}[1]{Mod. Phys. Lett. {\bf #1}} 
\newcommand{\ptp}[1]{Prog. Theor. Phys. {\bf #1}} 
\newcommand{\arns}[1]{Ann. Rev. Nucl. Sci. {\bf #1}}
\newcommand{\anfis}[1]{An. F\'\i s. {\bf #1}}
\makeatletter
\setlength{\clubpenalty}{10000}
\setlength{\widowpenalty}{10000}
\setlength{\displaywidowpenalty}{10000}

\vbadness = 5000
\hbadness = 5000
\tolerance= 1000
\arraycolsep 2pt

\footnotesep 14pt

\if@twoside
\oddsidemargin -17pt \evensidemargin 00pt \marginparwidth 85pt
\else \oddsidemargin 00pt \evensidemargin 00pt
\fi
\topmargin 00pt \headheight 00pt \headsep 00pt
\footheight 12pt \footskip 30pt
\textheight 232mm \textwidth 160mm

\let\@eqnsel = \hfil

\expandafter\ifx\csname mathrm\endcsname\relax\def\mathrm#1{{\rm #1}}\fi
\@ifundefined{mathrm}{\def\mathrm#1{{\rm #1}}}{\relax}

\makeatother

\unitlength1cm
\textheight 233mm

\begin{frontmatter}
\title {{\bf Parameters' domain in three flavour \\
 neutrino oscillations}}

\author{ J. Gluza}
and \author{M. Zra\l ek}

\address{ Department of Field Theory and Particle Physics, Institute 
of Physics, \\
University of
Silesia, Uniwersytecka 4, PL-40-007 Katowice, Poland}

\begin{abstract}
We consider analytically the domain of the three mixing angles 
$\Theta_{ij}$ and the CP phase $\delta$ for three flavour 
neutrino oscillations both in vacuum and matter. 
Similarly to the quark sector, it is necessary and sufficient 
to let all the mixing angles $\Theta_{12},\Theta_{13},\Theta_{23}$ and $\delta$ be in 
the range $\langle 0,\frac{\pi}{2} \rangle$ and $0 \leq \delta < 2 \pi$, respectively. 
To exploit the full range of $\delta$ will be important in future when more precise
fits are possible, even without CP violation measurements. With the above
assumption on the angles we can restrict ourselves to the natural order of
masses $m_1<m_2<m_3$. Considerations of the mass schemes with some
negative $\delta m^2$'s, though for some reasons useful, 
 are  not necessary from the point of view of neutrino oscillation
parametrization and cause  double counting only.
These conclusions are independent of matter effects.
\end{abstract}
\end{frontmatter}

\section{Introduction}

Three flavour neutrino oscillations are considered as a reliable mechanism to explain atmospheric and
solar neutrino anomalies. 
The neutrino flavour eigenstates
$\nu_\alpha=(\nu_e,\nu_\mu,\nu_\tau)$ are assumed to be combinations of 
mass eigenstates $\nu_i=(\nu_1,\nu_2,\nu_3)$
\begin{equation}
\nu_\alpha= \sum\limits_{i=1}^3 U_{\alpha i} \nu_i .
\end{equation}

Various parametrizations of the mixing matrix U are possible for Dirac and Majorana neutrinos. All of them use three mixing angles $\Theta_{ij}$
({ij}={12,13,23}) and one (Dirac) or three (Majorana) CP 
phases.  As the neutrino oscillation experiments are not sensitive 
to Majorana CP phases the same mixing matrix U as in the
quark sector \cite{pdg} is adopted (see e.g. \cite{fr} for discussion of various parametrizations) 

\begin{equation}
U=\left( 
\begin{array}{ccc}
c_{12}c_{13} & s_{12}c_{13} & s_{13}e^{-i \delta} \\ 
-s_{12}c_{23}-c_{12}s_{23}s_{13}e^{i \delta} & c_{12}c_{23}-s_{12}s_{23}s_{13}e^{i\delta} & s_{23}c_{13} \\ 
s_{12}s_{23}-c_{12}c_{23}s_{13}e^{i \delta} & -c_{12}s_{23}-s_{12}c_{23}s_{13}e^{i\delta} & c_{23}c_{13} 
\end{array}
\right) 
\label{u}
\end{equation}
where $c(s)_{ij} \equiv \cos(\sin){\Theta_{ij}}$.

The mixing angles $\Theta_{ij}$ can be defined to lie in the first
quadrant by appropriately adjusting the neutrino and charged lepton
phases, analogously to the quark sector \cite{leu}. To exhaust the full parameter
space the CP phase  $\delta$ must be taken in the range
$0 \leq \delta \leq 2 \pi$. There is only one important difference
between quark and neutrino sectors: alignment of absolute neutrino
masses is unknown and, among others, normal and inverse neutrino mass hierarchy
schemes are considered. It is also true that neutrino  
oscillations in vacuum will never be  able
to distinguish these two schemes. The argument is that in  vacuum, 
without CP violation measurements, the oscillation probability depends
only on $\sin^2{\left( \delta m_{ij}^2\frac{L}{4E} \right)}$, and 
the sign of $\delta m_{ij}^2$, which decide about the mass scheme, is 
unmeasurable. Neutrino oscillations in matter would only give the chance
to measure the sign of $\delta m_{ij}^2$. We would like to clarify 
the notion of using  $\delta m_{ij}^2$ signs for
neutrinos mixing  parametrization \cite{var}.
We find
the full domain of the three mixing angles  $\Theta_{ij}$
and $\delta$ phase in the mixing matrix U in matter. It appears that 
parameter space in the matter case is exactly the same as in vacuum

\begin{eqnarray}
0 \leq &  \Theta_{ij} & \leq \frac{\pi}{2}, \nonumber \\
0 \leq & \delta &  < 2 \pi.
\label{space}
\end{eqnarray}

In addition, in matter,
it is not necessary to consider mass schemes with different
mass arrangements $\delta m_{ij}^2=\pm |\delta {m_{ij}}^2|$.
With the full range of parameters, it is enough to include
only the ``canonical'' order of masses $(m_1 <m_2 <m_3 )$. 
All other mass schemes with negative  $\delta m_{ij}^2$
are equivalent to  that with
$\delta m_{21}^2>0$ and $\delta m_{32}^2>0$ 
and a different region in the  parameter space of Eq.~\ref{space}.
Finally we argue that exploring $\delta$ in its full range will be 
important in future experiments and not necessarily in 
connection with 
the explicit measurements of CP violation effects.

In the next Section, in a simple analysis we find the domain of
parameters for neutrino oscillation in vacuum. Even if the range
Eq.~\ref{space}  is well known from the 
quark sector,
we discuss it as it is a suitable introduction to understand the more
complicated case of neutrino oscillations in matter. This is presented
in Section 3. Finally, the  conclusions are gathered in Section 4.

\section{Parameter space for neutrino oscillations in vacuum}

The vacuum neutrino flavour oscillation probability for an initially
produced $\nu_\alpha$ with an energy E converted into detected $\nu_\beta$
after traveling a distance L is given by

\begin{equation}
P_{\nu_\alpha \to \nu_\beta} = 
\delta_{\alpha \beta}- 4 \sum\limits_{a>b} R^{ab}_{\alpha \beta}
\sin^2{\Delta_{ab}}- Y \sum\limits_{\gamma} \epsilon_{
\alpha \beta \gamma},
\label{prob}
\end{equation}

where

\begin{eqnarray}
R_{\alpha \beta}^{ab}&=&Re \left[ W_{\alpha \beta}^{ab} \right], \\
Y&=& 8 Im \left[ W_{e \mu}^{12} \right]  
\sin{\Delta_{21}} \sin{\Delta_{31}} \sin{\Delta_{32}}, 
\label{jar} \\
\Delta_{ab}&=&1.27 \delta m^2_{ab} [eV] \frac{L [km]}{E [GeV]}, 
\end{eqnarray}

and 

\begin{eqnarray*}
\delta m^2_{ab} &=&m_a^2-m_b^2,\;\;\;\;
W_{\alpha \beta}^{ab}=U_{\alpha a} U_{\beta b} U_{\alpha b}^{\ast}  U_{\beta a}^{\ast}.
\end{eqnarray*}

We can see that the mixing matrix elements $U_{\alpha a}$ enter the
oscillation probability by $W_{\alpha \beta}^{ab}$ tensors which are
invariant under the phase transformation

\begin{equation}
U_{\gamma c} \to e^{-i \delta_\gamma} U_{\gamma c} e^{i \eta_c}.
\end{equation}

Freedom of this transformation can be used to show that all mixing angles
$\Theta_{ij}$ originally belonging to the interval $\langle 0,2 \pi )$ can
be mapped onto the first quadrant $\Theta_{ij} \in \langle 0,\frac{\pi}{2} \rangle$.
As the  real and the imaginary parts of the phase factor $e^{i \delta}$ are
allowed to change  sign, 
the appropriate interval for $\delta$ is 
$ \langle 0,2 \pi)$. 

Now we show  in a
 way which will be useful in the more complicated case of neutrino
oscillations in matter, that in fact 
the domain from Eq.~\ref{space} covers the full 
parameter space of possible neutrino transitions. 
First of all, from unitarity of the U matrix
follows that all $R_{\alpha \beta}^{ab}$ tensors can be expressed
by squares of moduli of the U matrix elements

\begin{equation}
R_{\alpha \beta}^{ab}=\frac{1}{2} \left( |U_{\gamma a}|^2
 |U_{\gamma b}|^2- |U_{\alpha a}|^2 |U_{\alpha b}|^2-
 |U_{\beta a}|^2 |U_{\beta b}|^2 \right),
\label{r1}
\end{equation}
for $\alpha \neq \beta \neq \gamma, \;\;a \neq b $, and
\begin{equation}
R_{\alpha \alpha}^{ab}=|U_{\alpha a}|^2 |U_{\alpha b}|^2,\;\;
R_{\alpha \beta}^{aa}=|U_{\alpha a}|^2 |U_{\beta a}|^2,\;\;
\label{r2}
\end{equation}
otherwise.

The $|U_{e a}|^2$ 
for a=1,2,3 and $|U_{\alpha 3}|^2$ for
$\alpha=\mu,\tau$ depend only on sine and cosine squares
of $\Theta_{ij}$ and do not feel the transformations among
the four quadrants. Only $|U_{\alpha a}|^2$'s for  $\alpha=\mu,\tau$
and a=1,2 depend linearly on sines and cosines of  $\Theta_{ij}$ 
angles, namely

\begin{equation}
|U_{\alpha a}|^2=K_{\alpha a} \pm S,
\label{k}
\end{equation}

where $K_{\alpha a}$ are still functions of 
$\cos^2{\Theta_{ij}}$ and $\sin^2{\Theta_{ij}}$, but

\begin{eqnarray}
S&=&\frac{1}{2} F \cos{\delta}, \label{s} \\
&& \nonumber \\
F&=& \sin{2\Theta_{12}} \sin{2\Theta_{23}} 
 \sin{\Theta_{13}}. \label{f}
\end{eqnarray}

Exactly the same factor F appears in the Jarlskog invariant
(Eq.~\ref{jar}) \cite{jar}

\begin{equation}
J \equiv Im[W_{e \mu}^{12}]=\frac{1}{4} F  \cos^2{\Theta_{13}}
\sin{\delta}. \label{j}
\end{equation}

Only the F factor is sensitive to the change of sign
when the angles $\Theta_{ij}$ are mapped from the full domain
$\langle 0, 2 \pi )$ to the final range $\langle 0, \frac{\pi}{2} \rangle$ 
 ($n_{ij}=0,\ldots,3$)

\begin{eqnarray}
&&{\left. F\left( \Theta_{12}+n_{12} \frac{\pi}{2};
\Theta_{13}+n_{13} \frac{\pi}{2};
\Theta_{23}+n_{23} \frac{\pi}{2} \right) \right|}_{0 
\leq \Theta_{ij} \leq \frac{\pi}{2}} \nonumber \\
\;\;\;\;\;\;\;&=& (-)^{n_{12}+n_{23}} f(n_{13})
F\left( \Theta_{12};
\Theta_{13}; \Theta_{23} \right) \label{sign}
\end{eqnarray}
where
$$
f(n_{13})=\left\{ \matrix{+1\;\;\mbox{\rm for}\;\;n_{13}=0,1 \cr
                         -1\;\;\mbox{\rm for}\;\;n_{13}=2,3}.
\right.
$$ 

In order to compensate the possible change of signs in Eq.~\ref{sign}
other factors in Eqs.~\ref{jar},\ref{j} and  Eq.~\ref{s}
must have the freedom to change  sign.
The only possible choices are the CP phase $\delta$ and 
the combination $\Delta$ in Eq.~\ref{jar} defined as

\begin{equation}
\Delta=\sin{\Delta_{21}} \sin{\Delta_{31}} \sin{\Delta_{32}}. 
\label{delta}
\end{equation}

There are two possibilities. 
\begin{itemize}
\item
If $\delta \in \langle 0, 2 \pi )$ then a
change of sign by $\sin{\delta}$ in Eq.~\ref{j} and 
 $\cos{\delta}$ in Eq.~\ref{s} compensates the
 sign in Eq.~\ref{sign}.
In this case the order of masses can be kept 
canonical, $m_1<m_2<m_3$.
\item
  If $\delta \in \langle 0,  \pi )$ then  $\cos{\delta}$ is able to compensate 
the  sign in Eq.~\ref{s}, but  $\sin{\delta}>0$, so $\Delta$
must be used in the CP violating Y quantity (Eq.~\ref{jar}). 
In such a case, schemes with $\Delta>0$ ([123],[231],[312])
are distinguishable from schemes with $\Delta<0$ ([132],[321],[213])
in oscillation appearance experiments (see Fig.~\ref{ryssch} for notation). 
\end{itemize} 

We see that the chosen  $\Theta_{ij}$ and $\delta$ angles given in Eq.~\ref{space}
exhaust the full parameter space. We can bind the CP violating phase to the smaller
range 
\begin{equation}
(0 \leq \delta \leq \pi )
\label{smal}
\end{equation}

and in the same time distinguish the neutrino mass schemes with 
$\Delta>0$ (cyclic mass permutations
from the canonical case) 
from $\Delta<0$ cases (non-cyclic mass permutations of the canonical scheme).
It is impossible to disentangle schemes inside these two groups. 
Therefore an approach with the canonical order of masses
is clearer from the point of view of neutrino oscillation
parametrization: 
a point (region) in the parameter space of Eq.~\ref{space} 
determines the scheme of masses and the mixture of the weak states
in an unambiguously way. We will turn back to the  interpretation of $\delta m_{ij}^2$ 
signs in the next section.

\begin{figure}
\epsfig{figure=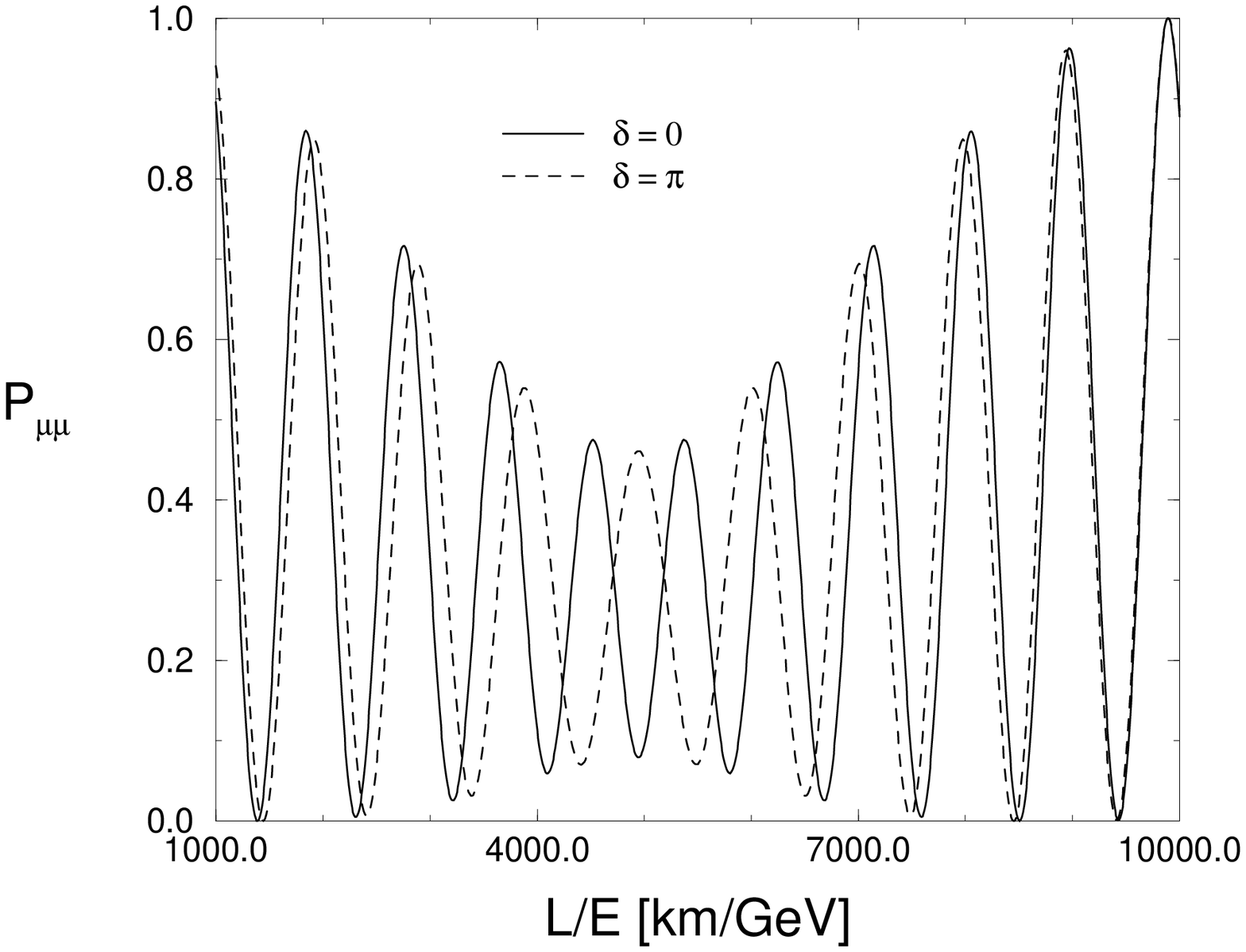, height=5 in}
\caption{The effect of the CP-violating phase in the disappearance $\nu_\mu \to \nu_\mu$
transition  (Eq.~\ref{pmm}).}
\label{vac}
\end{figure}

Presently, as statistical errors are large any  subleading effects in neutrino oscillations are neglected and
 experimental data for neutrino (disappearance) oscillations are fitted by the formula where only sine and/or cosine squares of the 
$\Theta_{ij}$ mixing angles appear. Therefore  we do not have to explore the full parameter space in Eq.~\ref{space}.
However, if the future precision improves and subleading effects 
are measured then it may be necessary to do it.
Now we show an example where taking into account the $\delta$ phase is important
even if CP violation is not measured. Let us consider atmospheric $\nu_\mu \to \nu_\mu$ disappearance probability
in vacuum

\begin{eqnarray}
P_{\mu \mu} &=&
1 \nonumber \\
&-&4 \left[ (K_{\mu 2} K_{\mu 1} -S^2) \sin^2{\Delta_{21}} \right. 
+ \left. \left| U_{\mu 3} \right|^2 ( K_{\mu 1} \sin^2{\Delta_{31}}
-K_{\mu 2} \sin^2{\Delta_{32}}) \right]  \nonumber \\
&-& 4  S \left\{ \left( K_{\mu 2}- K_{\mu 1} \right)
 \sin^2{\Delta_{21}} \right. 
+ \left. \left| U_{\mu 3} \right|^2
\left( \sin^2{\Delta_{31}}
-\sin^2{\Delta_{32}} \right) \right\}
\label{pmm}
\end{eqnarray}

where $K_{\mu i}$ (i=1,2) are defined in Eq.~\ref{k}. We can see that there is  a part proportional 
to S which exists only if $\Delta_{21} \neq 0 \Leftrightarrow \Delta_{31} \neq \Delta_{32}$.
In Fig.~\ref{vac} $P_{\mu \mu}$ as function of $L/E$ is given for 
$\delta m^2_{21} \equiv \delta m^2_{sol}=2.5 \cdot 10^{-4}\;eV^2$, 
$\delta m^2_{31} \equiv \delta m^2_{atm}=2.5 \cdot 10^{-3}\;eV^2$,
$\Theta_{23}=\Theta_{12}=\frac{\pi}{2}$ and
$\Theta_{13}=0.2$. $\delta$ is taken to be 0 and $\pi$.
The difference between $\delta=0$ and $\delta=\pi$ cases can be easily seen.
This difference diminishes with decreasing  $ \delta m^2_{sol}$.
Taking into account some new results where exploration of large values of
$ \delta m^2_{sol}$ (even up to a scale of $\delta m^2_{atm}$) is
discussed seriously \cite{lisi},\cite{str},\cite{sch},\cite{shr}  this effect should be keep in mind when a precise, global
analysis of oscillation data is undertaken, especially with incoming neutrino factory physics.
Let us note, that only $\delta=0$ is used presently. Usually it is assumed, that in 
the CP conservation case it is allowed to take $\delta=0$ and
$0 \leq \Theta_{ij} \leq \frac{ \pi}{2}$.
Surprisingly it is not true. 
If CP is conserved, then the discussion given above
implies that mapping the full range 
$0 \leq \Theta_{ij} < 2 \pi$ onto  $0 \leq \Theta_{ij} \leq \frac{ \pi}{2}$
requires the term $\cos{\delta}$ in Eq.~\ref{s}
to have two discrete values $\pm 1$.

\section{Parameter space for neutrino oscillations in matter}

The probability  $P_{\nu_\alpha \to \nu_\beta}^m$
of neutrino oscillations in matter of
density $N_e$ is given by the
vacuum  formula (Eq.~\ref{prob}) with modified $U_{\alpha a}, \Delta_{ab}$ and J \cite{xing}

\begin{equation}
P_{\nu_\alpha \to \nu_\beta}^m = 
\delta_{\alpha \beta}- 4 \sum\limits_{a>b} 
Re[W^{(m)ab}_{\alpha \beta}]
\sin^2{\Delta_{ab}^m}- Y^m \sum\limits_{\gamma} \epsilon_{
\alpha \beta \gamma},
\label{probm}
\end{equation}
where

\begin{eqnarray}
W_{\alpha \beta}^{(m)ab}&=&U_{\alpha a}^m U_{\beta b}^m U_{\alpha b}^{m \ast}
U_{\beta a}^{m \ast}, \\
&&  \nonumber \\
U^{\rm m}_{\alpha a} & = & \frac{N_a}{D_a} U_{\alpha a} 
+ \frac{A}{D_a} U_{e a} \left [ \left (\lambda_a^2 - m^2_b \right )
U^*_{e c} U_{\alpha c} 
+ \left (\lambda_a^2 - m^2_c \right )
U^*_{e b} U_{\alpha b} \right ] , \\
&&\mbox{\rm with}  \;\; a \neq b \neq c, \nonumber \\
&&  \nonumber \\
\Delta_{ab}^m&=& 1.27 \frac{(\lambda_a^2 - \lambda_b^2) [eV]^2  L [km] }{E [GeV] },  \\
&&  \nonumber \\
Y^m&=&8J^m \sin{\Delta_{21}^m} \sin{\Delta_{31}^m} 
\sin{\Delta_{32}^m}, \label{ym} \\
&&  \nonumber \\
J^m&=&  J \frac{\left( \lambda_1^2-m_2^2 \right) 
\left( \lambda_1^2-m_3^2 \right) \left( \lambda_2^2-m_1^2 \right) 
\left( \lambda_2^2-m_2^2 \right)}{D_1^2D_2^2} \label{jm} \nonumber \\
&\times &\left\{ N_1N_2-N_2A \left(  \lambda_1^2-m_2^2 \right) 
|U_{e1}|^2-N_1A \left(  \lambda_2^2-m_1^2 \right) 
|U_{e2}|^2 \right. \nonumber  \\
&+&  A^2 |U_{e1}|^2 |U_{e2}|^2 \left[
\left(  \lambda_2^2-m_1^2 \right) 
\left(  \lambda_1^2-m_3^2 \right)+
\left(  \lambda_2^2-m_3^2 \right)
\left(  \lambda_1^2-m_2^2 \right) \right. \nonumber \\
&-& \left. \left.
\left(  \lambda_2^2-m_3^2 \right)
\left(  \lambda_1^2-m_3^2 \right) \right] \right\}, \\    
&&  \nonumber \\ 
N_a & = & \left (\lambda_a^2 - m^2_b \right ) \left (\lambda_a^2 - m^2_c \right )
- A \left [\left (\lambda_a^2 - m^2_b \right ) |U_{e c}|^2  
+ \left (\lambda_a^2 - m^2_c \right ) |U_{e b}|^2 \right ] ,
\nonumber \\
&& \\
D^2_a & = & N^2_a + A^2 |U_{e a}|^2 \left [ 
\left (\lambda_a^2 - m^2_b \right )^2 |U_{e c}|^2  
+ \left (\lambda_a^2 - m^2_c \right )^2 |U_{e b}|^2 \right ] .
\end{eqnarray}

$\lambda_a^2$ denote the effective  mass squares of neutrinos 
in matter and follow from diagonalization
of an effective Hamiltonian  

\begin{equation}
{\cal H}_\nu = \frac{1}{2E} \left [ \left (
\matrix{
m^2_1     & 0      & 0 \cr
0         & m^2_2  & 0 \cr
0         & 0      & m^2_3 
} \right ) 
+ U^T \left ( 
\matrix{
A     & 0     & 0 \cr
0     & 0     & 0 \cr
0     & 0     & 0 \cr
} \right )  U^{\ast} \right ] .
\label{ham}
\end{equation}

$m_a$ (a=1,2,3) are neutrino masses and $A=2 \sqrt{2} E G_F N_e$.
Using the Cardano formula we get

\begin{eqnarray}
\lambda_1^2 & = & - \frac{a_2}{3} - \frac{1}{3} p \left( \cos{\phi} + \sqrt{3}  \sin{\phi} \right), \nonumber \\ 
\lambda_2^2 & = & - \frac{a_2}{3} - \frac{1}{3} p \left( \cos{\phi} - \sqrt{3}  \sin{\phi} \right), \label{lam} \\ 
\lambda_3^2 & = & - \frac{a_2}{3} + \frac{2}{3} p  \cos{\phi}, \nonumber 
\end{eqnarray}

where 

\begin{eqnarray}
p&=&\sqrt{a_2^2-3 a_1}, \nonumber \\
\phi&=&\frac{1}{3} \arccos \left[  - \frac{1}{p^3} \left(
a_2^3-\frac{9}{2} a_1 a_2+\frac{27}{2} a_0 \right) \right] ,
\end{eqnarray}

and 

\begin{eqnarray}
a_0&=&-m_1^2m_2^2m_3^2-A 
 \left[ m_1^2m_3^2|U_{e2}|^2 
+  m_1^2m_2^2|U_{e3}|^2
+ m_2^2m_3^2 |U_{e1}|^2 \right] , \nonumber \\ 
&& \nonumber \\
a_1&=&m_2^2m_3^2+m_1^2m_2^2+m_1^2m_3^2 \label{a} \\
& +& A \left[ m_1^2 (1-|U_{e1}|^2) 
 +  m_2^2 (1-|U_{e2}|^2)
+ m_3^2 (1-|U_{e3}|^2) \right] , \nonumber \\
&& \nonumber \\
a_2&=&-(m_1^2+m_2^2+m_3^2+A) . \nonumber
\end{eqnarray}

Now we can proceed as in the vacuum case and consider $P_{\nu_\alpha \to \nu_\beta}^m \left(  
\Theta_{ij}, \delta, \delta m_{ij}^2 \right)$ in the full range of parameters. Similarly to the vacuum case 
the real parts of $ W_{\alpha \beta}^{(m)ab}$  depend on $|U^m_{\delta c}|^2$. These 
subsequently depend on vacuum mixing matrix elements

\begin{eqnarray}
|U^m_{\alpha a}|^2 &=& \frac{N_a^2}{D_a^2} |U_{\alpha a}|^2+ 2\frac{A N_a}{D_a^2} \left\{
\left( \lambda_a^2-m_b^2 \right) R^{ac}_{e \alpha} + \left( \lambda_a^2-m_c^2 \right) R^{ab}_{e \alpha}
\right\} \\
&+& \frac{A^2}{D_a^2} |U_{e \alpha}|^2 \left\{ \left( \lambda_a^2-m_b^2 \right)^2 |U_{e c}|^2  |U_{\alpha c}|^2
+ \left( \lambda_a^2-m_c^2 \right)^2 |U_{e b}|^2  |U_{\alpha b}|^2  \right. \nonumber \\
&+& \left.
2  \left( \lambda_a^2-m_b^2 \right) \left(  \lambda_a^2-m_b^2 \right) R^{bc}_{e \alpha} \right\} . \nonumber 
\end{eqnarray}

We can see that the mixing angles appear in the  squared moduli $|U_{\gamma c}|^2$ and inside the R tensors
 (Eqs.~\ref{r1},\ref{r2}) which also depend on  $|U_{\gamma c}|^2$. So, as in the vacuum case, 
 when the full domain $\langle 0, 2 \pi )$ is mapped onto $\langle 0, \frac{\pi}{2} \rangle$,
non trivial signs appear only in the F factor (Eq.~\ref{f}).
Thus again, the  change of signs can be compensated by $\cos \delta$ in Eq.~\ref{s}.

In the CP violating part (Eq.~\ref{ym}) the vacuum mixing angles are found in the  squared moduli of
$|U_{ei}|^2$ and J (see Eq.~\ref{jm}). Again, only the F factor in J (Eqs.~\ref{f},\ref{j}) 
changes sign if the $\Theta_{ij}$
angles are reduced to the first quadrant. 
If  $\delta \in \langle 0, 2 \pi )$ then $\sin \delta$ term  in Eq.~\ref{j}
 is able to compensate the change of sign
in F. The mixing angles are also present  in the effective neutrino masses $\lambda_a$ (Eq.~\ref{lam}). However,  only 
$|U_{ei}|^2$ elements appear (Eq.~\ref{a}) and angles can be reduced to the first quadrant without changing  $\lambda_a$.
In this way we have proved that the domains of parameters for neutrino oscillations in matter and vacuum are the same
(Eq.~\ref{space}).

\begin{figure}
\epsfig{figure=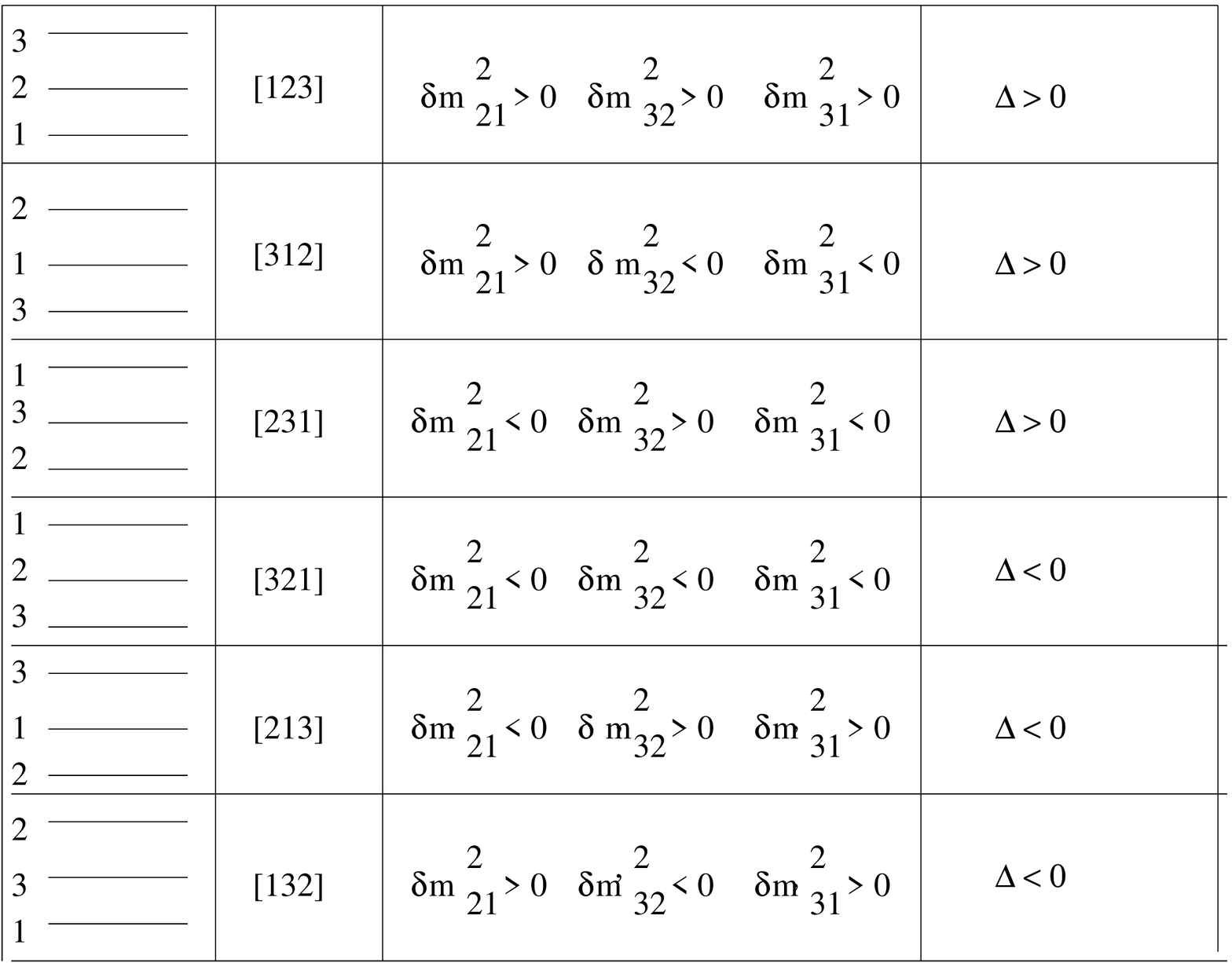, height=5 in}
\caption{Possible configurations of neutrino masses. The first one is called canonical.
The first two ([123],[312]) are usually discussed in literature. All schemes are completely equivalent since
marking neutrinos with numbers has no physical meaning. It is not important how a neutrino
with  number ``i'' couples to the $\alpha$ flavour. It is only meaningful  how neutrinos of different masses couple 
to the $\alpha$ flavour.}
\label{ryssch}
\end{figure}

Now we would like to answer  the question, whether introducing permutations of masses to the canonical scheme [123]
(see Fig.~\ref{ryssch}) is able to reduce the parameter space 
both for $\Theta_{ij}$ and $\delta$ in  Eq.~\ref{space}. 
Such an approach to the  $\Theta_{ij}$ angles was common  before
the ``dark side'' era \cite{dark}. Statements have also appeared
that $\delta \in ( 0, 2 \pi \rangle$ can be shrunk to half of
this region when negative signs of $\delta m_{ij}^2$ are 
taken into account.

Let us start our considerations from the  $\Theta_{ij}$ angles.
In the case of two flavour neutrino oscillations in vacuum the transition probability depends only on 
$\sin^2{2 \Theta} \sin^2 \left[ \delta m^2 \frac{L}{4E} \right]$. Then it is possible to limit the range of
mixing angles to the first octant. The transition probability in matter depends on the combination \cite{comb}

\begin{equation}
\left( \frac{A}{\delta m^2} - \cos{2 \Theta} \right)^2 .
\end{equation}

The relative sign between  $\delta m^2$ and $  \cos{2 \Theta}$ is important, so two possibilities are considered
\begin{eqnarray*}
&& \delta m^2 > 0 \;\;\; \mbox{\rm and}\;\;  0< \Theta \ < \frac{\pi}{2}  \\
\mbox{\rm or} && \\
&& \delta m^2 = \pm |  \delta m^2  | \;\;\;  \mbox{\rm and}\;\;  0< \Theta \ < \frac{\pi}{4}.
\end{eqnarray*}

In the case of  three flavour neutrino oscillations it is impossible to limit the range of $\Theta_{ij}$ angles in this way,
even in vacuum. Transition probabilities (Eq.~\ref{prob}) depend not only on  the product 
$\sin^2 \Theta_{ij} \cos^2 \Theta_{ij}$ 
but also on  $\sin^2 \Theta_{ij}$ and  
 $\cos^2 \Theta_{ij}$ separately\footnote
{This statement is general. In the approximation with one 
dominating $\delta m^2$ scale 
some transition probabilities depend only on $\sin^2{2 \Theta_{ij}}$. For example, the short-baseline reactor disappearance 
probability $P_{\nu_e \to \nu_e}=
1-\sin^2{\Theta_{13}} \sin^2{\Delta_{23}}$. 
However, this 
approximation seems to be questionable, even for present neutrino
data \cite{lisi} and quite probably a full theoretical framework without
neglecting some $\delta m^2$ should be used in future.}. 
Since it is impossible to shrink the range of the mixing angles $\Theta_{ij}$ in the case of vacuum oscillations,
the same will hold true for the matter case. 
In spite of that, various  schemes (Fig.~\ref{ryssch}) 
are considered. Let us show that in this way the same angles
are used twice when $0 < \Theta_{ij} < \frac{\pi}{2}$.

Neutrino oscillation formulae  
(Eqs.~\ref{lam}-\ref{a}) are symmetric under permutation of neutrinos.
Traditionally some scalar matrix $( {\bf {1}} \cdot const)$ 
is removed from the effective hamiltonian (Eq.~\ref{ham})
giving the same physical predictions. 
For example, if  $H_\nu$ is written in the form

\begin{equation}
M^2=
\left( \matrix{
1     & 0      & 0 \cr
0         & 1  & 0 \cr
0         & 0      & 1 
} \right ) m_1^2
+
\left( \matrix{
0     & 0      & 0 \cr
0         & \delta m^2_{21}  & 0 \cr
0         & 0      & \delta m^2_3 
} \right ) ,
\label{m2}
\end{equation}

then the matrix ${\bf 1} \cdot m_1^2$ can be absorbed
giving a common phase factor for all three neutrino
flavours. In such a case we diagonalize the hamiltonian $H_\nu$ where

\begin{equation}
m_1^2 \to 0, \;\; m_2^2 \to \delta m^2_{21} \;\; 
 m_3^2 \to \delta m^2_{31}.
\end{equation}

The new $a_i$ parameters derived from Eq.~\ref{a} 
are not symmetric under permutations of the masses anymore, 
they depend on $\delta m_{ij}^2$'s, namely

\begin{eqnarray}
a_0&=&-A 
  \delta m_{21}^2  \delta m_{31}^2 |U_{e1}|^2 ,
\nonumber \\
&& \nonumber \\
a_1&=&  \delta  m_{21}^2  \delta m_{31}^2  
+ A \left[  \delta m_{21}^2 (1-|U_{e2}|^2) 
 +  \delta  m_{31}^2 (1-|U_{e3}|^2 ) \right] , \nonumber \\
&& \nonumber \\
a_2&=&-\delta m_{21}^2- \delta m_{31}^2-A .   \label{abase}
\end{eqnarray}

Let us now calculate the eigenvalues for the case  
of negative $ \delta m_{3i}^2$, i=1,2, i.e. 
$ \delta m_{3i}^2=-
|  \delta m_{3i}^2 |$. We have

\begin{eqnarray}
a_0 ( -|  \delta m_{3i}^2 |) &=&A 
 | \delta m_{21}^2| |  \delta m_{31}^2| |U_{e1}|^2 ,
\nonumber \\
&& \nonumber \\
a_1 ( -|  \delta m_{3i}^2 |) &=& 
 - |\delta  m_{21}^2| |  \delta m_{31}^2 | 
 + A \left[  | \delta m_{21}^2 | (1-|U_{e2}|^2) 
 -  | \delta  m_{31}^2 | (1-|U_{e3}|^2 ) \right] ,\nonumber \\
&& \nonumber \\
a_2 ( -|  \delta m_{3i}^2 |) &=&-( | \delta m_{21}^2 | - | \delta m_{31}^2 | +A) .  \label{adel}
\end{eqnarray}

Using these new parameters $a_i ( -|  \delta m_{3i}^2 |)$ 
different  $\lambda_i^2$ eigenvalues  are obtained.
Are these new  $\lambda_i ( -|  \delta m_{3i}^2 |)$ eigenvalues
equal to the ``canonical'' $\lambda_i$ calculated at some other point
of the parameter space of Eq.~\ref{space}? 
To show that they are, let us
take the scheme [312]. This scheme (as any other in Fig.~\ref{ryssch}) is completely equivalent to the canonical one [123]. We have
only to change the names of particles $2 \to 3,\; 1 \to 2, \; 3 \to 1$ or more precisely replace 
$ U_{e1} \to U_{e2}, \;U_{e2} \to U_{e3}, \;
 U_{e3} \to U_{e1}, \; \delta m_{23}^2 \to  \delta m_{31}^2, \;
 \delta m_{21}^2 \to \delta m_{32}^2, \;
 \delta m_{13}^2 \to \delta m_{21}^2$. 

In   the scheme [312], as previously  we substract $m_1^2$ mass from the $M^2$ matrix. As now $m_3$ is the lightest mass,
we have to diagonalized  $H_\nu$ with the following replacements

\begin{equation}
m_1^2 \to 0,\;\;\;m_2^2 \to | \delta m_{21}^2 |,\;\;\;
m_3^2 \to - |\delta m_{31}^2|.
\end{equation}

The parameters $a_i$ which we get are exactly the same as given by Eq.~\ref{adel}. Similarly
we can check that any replacement $\delta m_{ij}^2 \to - \delta m_{ij}^2 $ in the canonical parameters $a_i ([123])$ is
equivalent to the others given by one of the six schemes in Fig.~\ref{ryssch}. In this way we have proved that changing the signs of 
$\delta m_{ij}^2 $ in the  canonical [123] eigenvalues is equivalent to evaluating  $\lambda_i^2$'s at some other point 
of the parameter space  Eq.~\ref{space}, schematically

\begin{equation}
\lambda_i^2 ( - | \delta m_{ij}^2 | ) \sim  \lambda_i^2 ( [ijk]  ).
\end{equation}

We can see that  using schemes with various permutations of masses does not confine the domain of the 
parameter space $\Theta_{ij}$ and causes
double counting only.
However, we can find a practical reason for introducing  
$\pm \delta m^2$'s. We have just shown that using various schemes 
is equivalent to using the [123] scheme with different values of
   $\Theta_{ij}$ angles in the parameter space. That is why we can
reverse the situation by fixing angles to the same physical 
situation, i.e.
 $\Theta_{12}$ can be connected with $\pm \delta m_{12}^2$ 
(oscillation of solar neutrinos),
$\Theta_{23}$ with   $\pm \delta m_{23}^2$ 
(oscillation of atmospheric neutrinos) and  $\Theta_{13}$ 
with reactor neutrino oscillations. 

Finally, let us consider the $\delta$ phase
in the case of matter neutrino oscillations. 
Can we bound it to the smaller range (Eq.~\ref{smal}) 
as in the vacuum case? 
There is a very elegant relationship between 
the universal CP-violating parameters $J^m$ and 
J in matter and in vacuum \cite{rel}

\begin{equation}
J^m (\lambda_2^2 - \lambda_1^2 )  (\lambda_3^2 - \lambda_1^2 ) (\lambda_3^2 - \lambda_2^2 ) =J
         (m_2^2 - m_1^2 )  (m_3^2 - m_1^2 ) (m_3^2 - m_2^2 ).
\end{equation}

From this relation  follows that the signs of $\delta m_{ij}^2$ and  $\delta \lambda_{ij}^2$ are correlated. If
 $\delta m_{ij}^2$ changes sign, the same happens to  $\delta \lambda_{ij}^2$.
We conclude that for neutrino oscillations in matter we have exactly the same situation as in the vacuum case. The basic domain of $\delta$ is
$\langle 0, 2 \pi )$ and it can be restricted to $\langle 0,  \pi \rangle $
and then the schemes with $\Delta > 0$ and $\Delta <0$ are distinguishable.

\section{Conclusions}
We have proved in an analytical way that the ranges of the mixing
angles $\Theta_{ij}$ and the CP violating phase $\delta$ are the
same for three flavour neutrino oscillations in vacuum and in matter:
$\Theta_{ij} \in \langle 0, \frac{\pi}{2} \rangle,\;\;
\delta \in \langle 0, 2 \pi ) $.
It means that probabilities for three flavour neutrino oscillations
can be described by points (more reliably by regions) 
in this parameters' domain without
using the signs of 
$\delta m_{ij}^2$ ($ \delta m^2_{ij}>0$,  $i>j$).
Contrary to the case of two neutrino oscillations in matter, 
the possibility
of two signs for each $\delta m^2_{ij}$ does not restrict further
the domain of the $\Theta_{ij}$ angles. Even though
the signs of  $\delta m_{ij}^2$'s are not needed, they are 
useful. Taking into account the signs of  $\delta m_{ij}^2$
we can fix angles  $\Theta_{ij}$ to a given scale of  
 $\delta m_{ij}^2$.  The range of the $\delta$ CP
phase can be confined to $ \delta \in \langle 0, \pi ) $ but
then, only sets of schemes with cyclic  ($\Delta > 0$) 
and odd  ($\Delta > 0$)  neutrino mass permutations are distinguishable
to each other. A simple example has been given that $\delta$ can
be important even for disappearance neutrino oscillation 
experiments.

\begin{ack}
We would like to thank M. Czakon for careful reading the manuscript
and useful discussion. 
This work was supported by Polish Committee for Scientific Research under 
Grants Nos. 2P03B04919 and 2P03B05418. 

\end{ack}

\newpage 

\newpage

\end{document}